# Digital Transformation and Corporate Financial Asset Allocation: Evidence from China

Yundan Guo[a], Han Liang[a], Li Shen[b,*]
a.China University of Political Science and Law, School of Business, 25 Xitucheng Road,Haidian District,Beijing,China,100088
b.School of Information and Electronics, Beijing Institute of Technology,5 South Zhongguancun Street,Beijing, 100081,China

# Abstract

Against the backdrop of rapid technological advancement and the deepening digital economy, this study examines the causal impact of digital transformation on corporate financial asset allocation in China. Using data from A-share listed companies from 2010 to 2022, we construct a firm-level digitalization index based on text analysis of annual reports and differentiate financial asset allocation into long-term and short-term dimensions. Employing fixed-effects models and a staggered difference-in-differences (DID) design, we find that digital transformation significantly promotes corporate financial asset allocation, with a more pronounced effect on short-term than long-term allocations. Mechanism analyses reveal that digitalization operates through dual channels: broadening investment avenues and enhancing information processing capabilities. Specifically, it enables firms to allocate long-term high-yield financial instruments, thereby optimizing the maturity structure of assets, while also improving information efficiency, curbing inefficient investments, and reallocating capital toward more productive financial assets. Heterogeneity analysis indicates that firms in non-eastern regions, state-owned enterprises, and larger firms are more responsive in short-term allocation, whereas eastern regions, non-state-owned enterprises, and small and medium-sized enterprises benefit more in long-term allocation. Our findings provide micro-level evidence and mechanistic insights into how digital transformation reshapes corporate financial decision-making, offering important implications for both policymakers and firms.

# Keywords

Digital transformation; Corporate financial asset allocation; Financialization; Staggered DID

# 1. Introduction

In today's complex and volatile global financial markets, optimal financial asset allocation is crucial for corporate performance. Sound allocation not only improves capital structure but also enhances operational efficiency and competitive advantage. However, many enterprises still face challenges due to limitations in traditional management methods, including inaccurate asset valuation, weak management systems, and insufficient technological application. First, market volatility and internal management deficiencies often lead to deviations between assessed and actual asset values. These inaccuracies may stem from delayed responses to market changes—such as demand shifts, technological innovations, or macroeconomic fluctuations—or from inadequate valuation methodologies. Inconsistent standards, incomplete information analysis, and lack of expertise further undermine valuation accuracy. Second, effective asset management requires clear policies, well-defined procedures, and robust monitoring mechanisms. Yet many firms exhibit poorly implemented policies, ambiguous execution processes, overlapping responsibilities, and ineffective oversight. Additionally, insufficient technological adoption inhibits data-driven decision-making. Persistent reliance on manual operations and subjective judgment reduces efficiency and increases error risk.

Digital transformation enables enterprises to achieve greater efficiency and transparency in asset management through advanced technologies such as big data and cloud computing. Big data analytics supports insightful assessment of asset utilization and cost-effectiveness, while cloud computing offers scalable processing power and real-time system accessibility. Furthermore, artificial intelligence applications like machine learning enhance predictive analytics for asset performance and maintenance, and improve decision-making through data-driven market forecasts. AI also strengthens risk management via real-time monitoring and early warning mechanisms. Ultimately, digital transformation serves as a strategic imperative for fostering innovation, addressing external challenges, and strengthening core competitiveness.

Corporate digital transformation refers to the process wherein enterprises leverage next-generation information technologies—such as the internet, IoT, big data, cloud computing, and artificial intelligence—to digitize and intelligentize products, services, operations, and management. With rapid advances in mobile internet, AI, and blockchain, digitalization has become an irreversible global trend. As a focal point for the deep integration of the digital and real economies, digital transformation is being vigorously accelerated worldwide. Countries are striving to promote comprehensive digital transformation to drive innovation in frontier technologies, achieve informatization, digitalization, and intelligentization in economic and social development, and secure a leading position in the new round of global competition, thereby continuously enhancing national comprehensive strength. In China, according to the "Digital China Development Report (2024)" released by the National Data Administration in 2025, the digital economy continues to expand, with the value-added of core industries accounting for approximately 10% of GDP. The scale of the digital economy is steadily increasing. Digital transformation has entered a phase of deep application: by the end of 2024, generative AI had attracted around 250 million users—meaning about one in every 5.6 Chinese uses AI tools. The industrial internet sector reached a value-added scale of RMB 5.01 trillion, a year-on-year increase of 6.5%, with over 230 top-tier smart factories cultivated[1]. Therefore, digital transformation has become a critical strategy for enterprises to overcome development bottlenecks and enhance international competitive advantage.。

Early studies also indicate that digital transformation enables firms to optimize resource allocation and enhance risk resilience. By leveraging large-scale data analytics, companies can improve risk prediction and control, and achieve automation and informatization in key business processes. This not only strengthens their adaptability to market conditions and risk tolerance, laying a solid foundation for sustainable operation, but also encourages greater engagement in high-risk innovative activities. A representative example is Alibaba, which has built an extensive digital platform ecosystem—including Taobao, Tmall, Alipay, and Alibaba Cloud—that offers integrated digital solutions through cross-platform collaboration. Its diversified business model spans B2B, B2C, C2C, e-commerce, cloud computing, digital media, and financial services, reducing operational risks and expanding revenue streams. Investments in AI, big data, cloud, and IoT technologies have further boosted operational efficiency, security, and customer experience. Through data-driven behavioral insights, Alibaba delivers personalized products and services, thereby increasing customer satisfaction and sales.

---

[1] National Data Administration. (2025, May). Digital China Development Report (2024) [Annual report]. from https://www.nda.gov.cn/

# 2.Literature Review

## 2.1Research on Corporate Financial Asset Allocation

### 2.1.1Definition of Corporate Financial Asset Allocation

There is no clearly established consensus on the definition of "corporate financial asset allocation." Early studies often used the proportion of financial assets in a firm's total investments as a direct measure of its financial asset allocation level. Stockhammer (2004) referred to the phenomenon by the increasing engagement of non-financial firms in financial activities, and measured the extent of such allocation by comparing financial expenses to cash flows[44]. As research evolved, however, scholars have argued that not all financial investments should be simplistically classified as financial asset allocation. It has been proposed that only when non-financial firms hold substantial financial assets with high degrees of virtualization and derive a significant portion of their profits from financial channels can they be considered as engaging in financial asset allocation[31]. In other words, a higher ratio of financial assets indicates a greater degree of financial asset allocation. Furthermore, firms continuously optimize cross-cycle resource allocation by adjusting the duration, risk exposure, and liquidity tiers of financial assets[9] (Krippner, 2011; Begenau & Salomao, 2023).

### 2.1.2Determinants of Corporate Financial Asset Allocation

The drivers of corporate financial asset allocation are a key research focus. Studies have explored micro-level factors such as investment returns, financing constraints, and corporate governance, as well as macro-level factors like economic policy.

1) Micro-level Perspectives

Corporate financial asset allocation represents a strategic investment behavior, which has been examined in prior research from perspectives such as investment returns, financing constraints, and corporate governance.

The high returns of financial assets serve as a primary driver. Categorized by maturity, financial assets are divided into long-term and short-term holdings. Firms often show a preference for highly liquid, short-term financial assets for speculative arbitrage to maintain flexibility (Seo et al., 2012)[42]. Furthermore, a declining return on physical investments widens the interest rate spread, which directly encourages corporate financialization (Demir, 2009)[19].

Financing constraints also contribute significantly to the over-allocation of financial assets. Firms facing stringent financing limitations, especially those under hard budget constraints, are more inclined to pursue financial asset returns to meet urgent funding needs (Demir and Ersan, 2017)[18]. Conversely, alleviating these constraints—for instance, through digital transformation—can help refocus corporate investments toward real economic activities (Smith, 2020)[43].

The objective of maximizing shareholder value and managerial self-interest also markedly influence financial asset allocation decisions. Equity incentives that tie executive compensation to stock performance may encourage greater allocation into financial assets to boost short-term share prices (Lazonick, 2010; Orhangazi, 2008)[32][38]. Similarly, compensation structures linked to current profits motivate executives to increase financial investments for immediate gains (Lin and Tomaskovic Devey, 2013)[33]. On the other hand, institutional investors, as key governance actors, can mitigate short-termism in financial allocation. Higher ownership by such investors helps align management focus with long-term value rather than short-term arbitrage (Aghion et al., 2013)[2]. Although a CEO's financial background may facilitate financial asset allocation, highly capable CEOs may conversely restrain excessive financial investments (Ishida and Kochiyama, 2021; Fabio et al., 2018)[29][21].

2) Macro-level Perspectives

Rising macroeconomic uncertainty prompts firms to increase holdings of liquid financial assets as a precaution against potential shocks, while reducing long-term investments (Bloom et al., 2007)[11]. The level of financialization typically exhibits pro-cyclical characteristics, with corporate financial asset allocation being more pronounced during economic booms than in recessions (Davis, 2018; Orhangazi, 2008)[17][39]. Moreover, economic policy uncertainty may shift corporate investment focus toward financial markets, crowding out investments in core business activities—a phenomenon particularly evident among private enterprises. Although industrial policies can stimulate investment enthusiasm, they may also suppress short-term value creation and compress room for physical investment, indirectly encouraging greater allocation to financial assets (Ahmed et al., 2020)[3].

## 2.2. Research on Digital Transformation

### 2.2.1 The Meaning of Digital Transformation

Prior to discussing the concept of digital transformation, it is essential to define digital technology. Previous research describes digital technology as a combination of information, computing, communication, and connectivity technologies that reshape organizational capabilities (Bharadwaj et al., 2013)[10]. A review of the literature on digital transformation reveals a lack of consensus regarding its precise definition. Integrating earlier perspectives, Vial (2019) characterizes digital transformation as a process that leverages digital technologies to trigger significant changes in the attributes of an entity, thereby enhancing its overall performance[47]. Verhoef et al. (2021) conceptualize it as a three-stage progression—digitization, digitalization, and digital transformation—in which digital transformation specifically refers to the use of digital technologies to develop new digital business models, enabling firms to capture greater value[46].

### 2.2.2 Measuring Digital Transformation

Existing studies measure digital transformation through multi-dimensional frameworks, primarily including dummy variables, composite indices, intangible assets ratios, textual analysis, and hybrid methods. The dummy variable approach assigns 1 or 0 based on the implementation of digital initiatives, commonly used in early technology adoption research (Brynjolfsson et al., 2021)[14]. To capture greater nuance, some studies develop multi-dimensional indices evaluating digital infrastructure, media usage, and transactions (Foster and Graham, 2017)[23]. Another strand uses the ratio of digital-related intangible assets—such as software and patents—to total assets to proxy transformation intensity (Haskel and Westlake, 2017; Babina et al., 2024)[27][4]. Textual analysis has become a mainstream method, constructing continuous measures based on keyword frequencies in corporate disclosures (Goldfarb and Tucker, 2019; Bui and Duong, 2025)[15][25]. Hybrid approaches combine textual, financial, and indicator-based data for comprehensive assessment.

### 2.2.3 Effects of Digital Transformation

Digital transformation exerts profound influences on firm operations and strategic decision-making by enhancing information processing capabilities, optimizing risk management, and restructuring organizational frameworks. The widespread adoption of digital technologies facilitates a shift toward networked and flatter organizational structures, enhancing inter-firm collaborative innovation and data sharing (Horvitz et al., 1988)[28]. Leveraging big data and artificial intelligence, management can acquire and integrate internal and external information in real time, improving resource allocation efficiency, enabling rapid responses to market changes, and enhancing the prediction and management of decision risks (Goldfarb & Tucker, 2019)[25]. Furthermore, digital transformation allows firms to simulate R&D processes and operational decisions in virtual environments, significantly reducing trial-and-error costs and operational uncertainties. Digital technologies empower enterprises to identify and assess risks in real time, supporting more precise strategic responses (Michael et al., 2018)[36]. These changes collectively enhance firms' adaptability and competitive advantage in complex market environments.

## 2.3 Research on the Impact of Digital Transformation on Corporate Financial Asset Allocation

Although the literature on the economic consequences of digital transformation has grown in recent years, relatively few studies directly address its impact on corporate financial asset allocation. Existing research has primarily focused on the real economic effects of digital transformation, such as its role in promoting innovation or enhancing productivity, while its implications for financial investment decisions remain underexplored. To address this gap, this subsection examines potential mediating mechanisms—including information processing capacity, risk management capabilities, and resource reallocation and innovation—drawing on relevant theoretical and empirical literature to establish a foundation for understanding how digital transformation reshapes corporate financial asset allocation strategies.

### 2.3.1 Theoretical Research

The adoption of digital technologies such as artificial intelligence and blockchain enables firms to fundamentally transform strategic thinking, optimize business processes, organizational structures, and management models, thereby enhancing both technological and managerial efficiency (Vial, 2019)[47]. Such technological applications strengthen firms' control over external information and sensitivity to uncertainty, expand the sample size and key factor identification in R&D evaluation, effectively reduce innovation trial-and-error costs and mitigate R&D risks, ultimately influencing financial decision-making.

From a theoretical perspective, digital transformation influences financial asset allocation through multiple mechanisms. First, it significantly enhances corporate information processing capacity. Digital technologies enable firms to collect, analyze, and interpret large volumes of internal and external data in real time, reducing information asymmetry and improving investment precision (Marion et al., 2021)[35]. For instance, big data analytics facilitates the identification of mispriced financial assets, while AI algorithms improve portfolio management and risk assessment (Begenau et al., 2018)[7].

Second, digital transformation strengthens risk management capabilities. Technologies such as blockchain and smart contracts provide transparent and secure mechanisms for executing and monitoring financial transactions, reducing counterparty risks and enhancing trust in financial engagements (Gomber et al., 2017)[26]. Moreover, digital platforms allow firms to simulate financial scenarios, conduct stress tests on portfolios, and dynamically adjust asset allocations in response to market fluctuations (Nambisan et al., 2019)[37]. These capabilities reduce uncertainty and encourage greater involvement in financial investments, particularly in longer-term and higher-yield instruments.

Additionally, digital transformation facilitates resource reallocation and innovation. By transcending traditional resource boundaries and enabling more flexible asset utilization, digital technologies help firms redeploy idle capital into more productive financial uses (Loebbecke & Picot, 2015)[34]. For example, IoT and cloud platforms enhance asset liquidity, promote the securitization of physical assets, and broaden the range of investable financial products (Balyuk & Davydenko, 2024)[5].

It is worth noting, however, that digital transformation may also amplify financial risks under certain conditions. Enhanced risk-taking capacity could lead to overly aggressive financial strategies, particularly in contexts with weak corporate governance (Forman & van Zeebroeck, 2018)[22]. Therefore, a nuanced empirical understanding remains essential to accurately assess how digital transformation shapes financial asset allocation.

### 2.3.2 Empirical Evidence

A growing body of empirical evidence in recent years has shed light on the profound impact of digital transformation on corporate financial resource allocation, particularly forming systematic conclusions regarding its role in enhancing risk-taking capacity and optimizing financing structures.

Digital transformation significantly elevates corporate risk-taking by improving information processing and decision-making capabilities. Begenau et al. (2018)[65], using textual measures of big data analytics adoption among U.S. firms, found that such technologies led to a higher allocation to high-yield risky assets and an average increase of 12% in risk-adjusted portfolio returns. Bloom et al. (2014)[12], analyzing survey data from European manufacturing firms, demonstrated that a one-standard-deviation increase in IT investment reduced management layers by 0.3 and accelerated decision-making by 18%, thereby strengthening firms' capacity to respond to and absorb market volatility.

In terms of financing structure, digital transformation reduces information asymmetry and future uncertainty, thereby weakening precautionary savings motives. Duchin et al. (2017)[20], using COMPUSTAT data, showed that firms with higher levels of digitalization held significantly less cash and had 4.7 percentage points more long-term debt in their capital structure, confirming that digital technologies mitigate precautionary motives and facilitate long-term financing. Balyuk & Davydenko (2024)[68], drawing on cross-country firm-bank matched data, found that firms using digital lending platforms were 23% more likely to secure long-term loans, with debt maturity extended by approximately 1.5 years, indicating that digital finance platforms effectively improve access to long-term debt.

Babina et al. (2024)[30] provided further supporting evidence: using machine learning to measure AI adoption among U.S. public firms, they identified a positive correlation between AI intensity and investment in top-percentile innovation projects. These firms also increased strategic financial asset holdings by an average of 2.4%, suggesting that digital transformation not only affects the scale but also optimizes the structure of financial investments toward long-term strategic goals. Additionally, Balyuk & Davydenko (2024)[68], analyzing investment portfolio data across 40 countries with panel fixed-effects models, reported that fintech adoption raised the Sharpe ratio by 0.18 on average, attributable to improved diversification and more accurate market timing facilitated by digital technologies.

## 2.4. Research Gaps and This Study's Contributions

Existing research has yielded valuable insights into financial asset allocation, digital transformation, and their interplay, providing important theoretical foundations and analytical frameworks for this study. However, significant research gaps remain that warrant further investigation. First, the literature shows inconsistent findings on how digital transformation affects financial investments, with some studies identifying a positive effect and others a negative one. These contradictions likely arise from variations in model design, measurement approaches, and sample composition—especially the insufficient distinction between short- and long-term financial assets. Second, most extant mechanistic research focuses on isolated pathways such as financing constraints or information efficiency, and fails to offer a holistic theoretical framework incorporating both internal and external mechanisms. Insufficient attention has been given to the interplay between internal elements like information processing, governance, and resource reallocation, and external factors such as financial market development, policy environment, and industry competition. As a result, the understanding of how digital transformation shapes corporate financial behavior remains incomplete.

In response to these limitations, the marginal contributions of this paper are threefold:

First, it introduces an innovative perspective by distinguishing between short- and long-term financial asset allocations and examining the heterogeneous effects of digital transformation through both macro- and micro-dimensions. This study not only considers macro-level contextual factors such as regional financial market development and industrial policy but also incorporates micro-level characteristics including ownership type, firm size, and governance structure. This approach reveals differential responses to digital shocks across institutional settings and firm traits, offering a contextualized understanding of the economic consequences of digital transformation and supporting evidence-based, targeted policy design.

Second, this research expands the mechanistic pathways by proposing and testing a "dual-channel" framework. On one hand, it examines how digital transformation alleviates financing constraints, broadens investment channels, and optimizes asset maturity structure through external capital market conditions. On the other hand, it investigates how digitalization enhances internal operational capacities—such as improving information transparency, curbing inefficient investments, and reallocating capital toward productive financial uses—thereby increasing the efficacy of financial asset allocation. This integrated framework offers new insights into the causal pathways between digitalization and corporate financial behavior.

Third, the study makes methodological improvements by constructing a multidimensional, continuous digital transformation index using textual analysis to overcome the limitations of binary or single-measure proxies. This approach allows more precise identification of differential effects on long- versus short-term asset allocation. Furthermore, the empirical strategy employs staggered difference-in-differences (DID), instrumental variable (IV) approaches, and Bootstrap testing to strengthen causal identification and inference robustness.

# 3.Mechanism Analysis

## 3.1Broadening of Investment Channels

Digital transformation reshapes the financial landscape and expands corporate investment channels, serving as a key driver of optimized financial asset allocation (Gomber et al., 2017)[26]. It facilitates a strategic shift in asset structure by enabling access to long-term instruments such as private equity and cross-border REITs, thereby transitioning portfolios from short-term liquidity management to long-term strategic investment. This is directly reflected in the rise of Fin2_Ratio, the proportion of long-term financial assets.

First, diversified product systems are transforming investment options. Blockchain technology has facilitated the emergence of long-term high-yield assets such as security token offerings and infrastructure REITs (Corbet et al., 2019)[16]. Tokenized private equity accounted for 37% of the $1.2 trillion global digital asset market in 2023 (CoinMarketCap), directly elevating Fin2_Ratio. Second, open banking and API technologies lower barriers to cross-border investment. Over 100 jurisdictions have implemented open banking frameworks, connecting firms to global capital markets via data sharing[6]. For example, the EU's PSD2 directive standardizes bank APIs, allowing access to over 200 European banks and improving cross-border payment efficiency by more than 40% (BIS, 2023), thus facilitating offshore long-term asset allocation. Third, AI algorithms improve precision in asset allocation. Machine learning identifies undervalued long-term assets like green bonds. Citi's platform processes 15TB daily, improving long-term allocation accuracy by 41% and raising Fin2_Ratio by 2.3 percentage points on average (Begenau et al., 2018)[7]. Digital platforms also democratize access, offering structured long-term products with minimum investments as low as $100 (Vallee & Zeng, 2019)[45]. A 10% increase in Fin2_Ratio raises the Sharpe ratio by 0.36 (Balyuk & Davydenko, 2024)[5], confirming enhanced risk-adjusted returns. In summary, digital transformation elevates Fin2_Ratio, forming a transmission mechanism of "channel expansion →

long-term holding → efficiency gains." Thus, we propose:
H1: Digital transformation improves the efficiency and risk-adjusted returns of financial asset allocation by increasing Fin2_Ratio.

## 3.2 Enhancement of Information Acquisition and Analysis Capability

Corporate digital transformation enhances financial resource allocation by improving information acquisition and analysis, optimizing asset allocation efficacy (Brynjolfsson & McAfee, 2014)[13]. Its core mechanism involves structural upgrades to the information ecosystem: blockchain and IoT enable integrated, real-time data collection across supply chains, consumer behavior, and macroeconomics, breaking down data silos (Gomber et al., 2017)[26]. Machine learning and AI further refine investment precision—J.P. Morgan (2024) used NLP and satellite imagery to boost risk-adjusted returns by 18%, while BlackRock's Aladdin reduced client drawdowns by 14% in the 2022 volatility (BlackRock, 2023). These advances lower information costs and improve predictive accuracy, curbing inefficient investments and shifting capital toward high-yield assets. Goldman Sachs' Marquee platform anticipated the 2023 banking crisis via sentiment analysis, avoiding $3.5 billion in losses and reallocating capital toward long-term equity investments—demonstrating a clear transmission channel from "information acquisition →efficiency optimization →capital reallocation." Thus, through a four-dimensional mechanism—expanding information breadth, deepening analytical insights, accelerating response speed, and reconstructing investment efficiency—digital transformation shifts financial asset allocation from experience-based to data-intelligence-driven. Accordingly, we propose:
H2: Digital transformation improves financial asset allocation efficiency by enhancing information processing and reducing inefficient investment.

## 4. Research Design

### 4.1 Model Specification

Building on a systematic review of the theoretical mechanisms linking digital transformation and financial asset allocation, this study draws on the fixed-effects approach applied by Acemoglu & Restrepo (2019)[1] and the measurement and determinant framework of corporate financial asset allocation developed by Begenau & Palazzo (2018)[8]. To mitigate endogeneity concerns, a progressive empirical strategy is constructed. It begins with a baseline fixed-effects model to preliminarily examine the correlation between corporate digital transformation and financial asset allocation:

$$Fin_{i,t} = \beta_0 + \beta_1 Digital_{i,t} + \gamma \mathbf{X}_{i,t} + \alpha_i + \delta_t + \varepsilon_{i,t} \quad (1)$$

The model controls for both firm fixed effects $\alpha_i$ and year fixed effects $\delta_t$, where $Digital_{i,t}$ is a continuous measure of corporate digital transformation, and $\mathbf{X}_{i,t}$ represents a set of time-varying controls including *Size, LEV, ListAge*, and other firm-level characteristics.

To address potential endogeneity concerns such as reverse causality and selection bias inherent in the baseline specification, this study further innovatively employs a multi-intensity, multi-period staggered Difference-in-Differences design as the core causal identification strategy:

$$Fin_{i,t} = \beta_0 + \beta_1 (Digital_i \times Post_{i,t}) + \gamma \mathbf{X}_{i,t} + \alpha_i + \delta_t + \varepsilon_{i,t} \quad (2)$$

This model addresses two major limitations of conventional Difference-in-Differences (DID) designs—binary treatment and simultaneous treatment timing—through two key innovations: First, the continuous digital transformation measure $Digital_i$ is transformed into a binary treatment variable, where firms with medium-to-high intensity are assigned a value of 1 and those with low intensity 0, capturing intensity heterogeneity. Second, a time-varying indicator $Post_{i,t}$ is introduced, which equals 1 from the year of digital transformation onward and 0 otherwise, accommodating the asynchronous timing of firm-level digital adoption. The coefficien $\beta_1$ on the interaction term captures the net causal effect of digital transformation on corporate financial asset allocation and is the main parameter of interest.

Variable definitions remain consistent across models: the outcome variable $Fin_{i,t}$ encompasses both short-term *fin1* and long-term *fin2* financial asset allocations. The control set $X_{i,t}$ follows the baseline model specifications. Standard errors are clustered at the firm level, and $\varepsilon_{i,t}$ denotes the idiosyncratic error term. This framework preserves the causal identification advantages of traditional DID while explicitly accounting for heterogeneous intensity and staggered adoption of digital transformation, offering a methodological innovation for estimating its impact on financial asset allocation.

## 4.2 Variable Definitions

## 4.2.1 Digital Transformation

Our study adopts the text analysis methodology following Wu et al. (2021)[49] to construct a firm-level digital transformation index. Specifically, Python-based web crawlers are used to collect annual reports of A-share listed companies. A comprehensive metric reflecting the extent of AI adoption is developed by identifying and counting the frequency of predefined keywords (as summarized in Table 1). Compared to dummy-variable approaches commonly used in earlier literature, this method better captures varying intensities of digital application and provides a improved data foundation for staggered Difference-in-Differences (DID) quasi-natural experiments.

The construction process involves four steps: First, core keywords related to digital transformation are identified based on national policy documents and academic literature. Second, environmentally relevant digital terms frequently appearing in annual reports are added to expand the initial lexicon. Third, following official reports on digital transformation in state-owned and private enterprises, keywords are categorized into five dimensions: "Digital Technology Application," "Digital Information Systems," "Intelligent Digital Management," "Digital Marketing Models," and "Digital Efficiency Enhancement." Finally, the total frequency of these keywords in each firm's annual report is calculated and natural-log transformed to form the digital transformation index. A higher value indicates a greater degree of digital transformation.

Table 1: Keyword Lexicon for Digital Transformation

| Dimension | Category Terms | High-Frequency Keywords |
|---|---|---|
| Digital Technology Applications | Data, Digital, Digitalization | Data Management, Data Mining, Data Network, Data Platform, Data Center, Data Science, Digital Control, Digital Technology, Digital Communication, Digital Network, Digital Intelligence, Digital Terminal, Digital Marketing, Digitalization, Big Data, Cloud Computing, Cloud IT, Cloud Ecosystem, Cloud Service, Cloud Platform, Blockchain, Internet of Things (IoT), Machine Learning, Artificial Intelligence (AI), Business Intelligence, Image Understanding, Investment Decision Support System, Intelligent Data Analysis, Intelligent Robot, Deep Learning, Semantic Search, Biometric Technology, Facial Recognition, Voice Recognition, Identity Verification, Autonomous Driving, Natural Language Processing (NLP), Text Mining, Data Visualization, Heterogeneous Data, Augmented Reality (AR), Mixed Reality (MR), Virtual Reality (VR), Stream Computing, Graph Computing, In-Memory Computing, Multi-Party Secure Computation, Neuromorphic Computing, Green Computing, Cognitive Computing, Converged Architecture, Massively Parallel Concurrency, Exabyte-Scale Storage, Cyber-Physical System (CPS), Digital Currency, |

| | | Distributed Computing, Differential Privacy Technology |
|---|---|---|
| Digital Information Systems | Information, Informatization, Networking | Information Sharing, Information Management, Information Integration, Information Software, Information System, Information Network, Information Terminal, Information Center, Informatization, Networking |
| Intelligent Digital Management | Intelligent, Automation, Numerical Control, Integration | Artificial Intelligence, Advanced Intelligence, Mobile Intelligence, Intelligent Control, Intelligent Terminal, Intelligent Mobility, Intelligent Management, Intelligent Logistics, Intelligent Warehousing, Intelligent Technology, Intelligent Equipment, Intelligent Production, Intelligent Connected System, Intelligent System, Automation, Automatic Control, Automatic Monitoring, Automatic Detection, Automatic Production, Numerical Control (NC), Integration |
| Digital Marketing Models | Internet, E-Commerce | Mobile Internet, E-Commerce, Mobile Payment, Third-Party Payment, NFC Payment, Smart Energy, B2B, B2C, C2B, C2C, O2O, Internet of Vehicles (IoV), Industrial Internet, Internet Solution, Internet Technology, Internet Thinking, Internet Initiative, Internet Business, Internet Mobility, Internet Application, Internet Marketing, Internet Strategy, Internet Platform, Internet Model, Internet Business Model, Internet Ecosystem, E-Business, E-Commerce, Internet, Internet Plus, Online-to-Offline (O2O) |
| Digital Efficiency Enhancement | Smart, Intelligent | Smart Agriculture, Intelligent Transportation, Intelligent Customer Service, Smart Culture & Tourism, Intelligent Environmental Protection, Smart Grid, Intelligent Marketing, Digital Marketing, Unmanned Retail, Internet Finance, Digital Finance, FinTech, Financial Technology, Quantitative Finance, Open Banking |

### 4.2.2 Financial Asset Allocation

To accurately identify the heterogeneous impact channels of digital transformation and avoid estimation biases caused by conflating different types of financial assets, this study adopts a strategy that distinguishes between long- and short-term financial asset allocations (Peng et al., 2018)[40]. The rationale is fundamental: long-term allocations (e.g., equity investments) reflect strategic positioning and rely on in-depth information analysis and risk management, whereas short-term allocations (e.g., trading financial assets) emphasize liquidity management and short-term arbitrage, being highly sensitive to real-time transactional efficiency—digital transformation affects these two types in fundamentally distinct ways.

Examining them separately not only reveals how digitalization reshapes corporate financial behavior through differentiated mechanisms (e.g., enabling strategic decision-making in the long term and reducing transaction costs in the short term), but also facilitates targeted risk warnings: long-term risks primarily relate to credit mismatch, while short-term risks are more tied to market volatility. This approach provides a stratified governance framework for both regulators and firms.

The measurements for corporate financial asset investments are constructed as follows:
Short-term financial assets (fin1) = (Monetary Funds + Financial Assets at Fair Value Through Profit or Loss) / Total Assets
Long-term financial assets (fin2) = (Net Interest Receivable + Net Dividends Receivable + Net Available-for-Sale Financial Assets + Net Held-to-Maturity Investments + Net Long-Term Equity Investments + Net Investment Properties) / Total Assets

### 4.2.3 Control Variables

To mitigate omitted variable bias and enhance model rigor, this study controls for key firm-level characteristics influencing financial asset allocation, following Freund et al. (2021)[24], including firm Size:log of total assets), LEV:liability-to-asset ratio, ROA:return on assets, TOP1:largest shareholder's ownership percentage, ListAge:log of listing years, TobinQ:market-to-book ratio, and Indep:share of independent directors. These variables account for differences in investment capacity, financial constraints, profitability, corporate governance, life-cycle stage, growth opportunities, and board monitoring intensity.

### 4.3 Data Sources

This study employs data from Chinese A-share listed companies spanning 2010 to 2022. Financial and stock market data are sourced from the Wind and CSMAR databases, while annual reports are obtained from the official websites of the Shanghai and Shenzhen stock exchanges. Data processing and analysis are conducted using Python and Stata. The sample is treated as follows: (1) exclude manufacturing firms that have been delisted or are under ST/*ST status; (2) remove observations with missing values in key variables; and (3) winsorize all continuous variables except explanatory and dummy variables at the 1st and 99th percentiles to mitigate the influence of outliers.

To ensure model robustness, extreme observations were prudently handled. Standardized residual analysis identified 20 influential observations exerting excessive leverage on the coefficient of the digital variable, likely due to measurement error or sampling bias, which could distort asymptotic efficiency.

Table 2 presents descriptive statistics. The average short-term financial asset ratio (fin1, 19.59%) substantially exceeds the long-term ratio (fin2, 5.55%), indicating a corporate preference for liquidity under financing constraints and uncertainty. Digital adoption remains low and uneven, suggesting early-stage technological penetration. With mean leverage at 42.72% and occasional debt overload, some firms face heightened default risks that may crowd out real investment. The average Tobin's Q of 2.07 reflects market overvaluation, potentially encouraging financial arbitrage. These patterns depict an economy marked by liquidity hoarding, slow digitalization, and coexisting debt risks and valuation distortions.

Table 2: Descriptive Statistics

| Variable | N | mean | sd | min | max |
| --- | --- | --- | --- | --- | --- |
| fin1 | 27,897 | 0.1959 | 0.1351 | 0.0007 | 0.9167 |
| fin2 | 27,897 | 0.0555 | 0.0800 | -0.0000 | 0.8379 |
| digital | 27,903 | 0.0122 | 0.0240 | 0.0001 | 0.4554 |
| Size | 27,903 | 22.2610 | 1.2860 | 17.4260 | 28.6365 |
| LEV | 27,903 | 0.4272 | 0.2332 | 0.0075 | 10.4953 |
| ROA | 27,903 | 0.0372 | 0.0926 | -1.8719 | 4.4890 |
| ListAge | 27,903 | 2.1370 | 0.7706 | 0.6931 | 3.4965 |
| TobinQ | 27,436 | 2.0669 | 1.4675 | 0.6245 | 45.9734 |
| TOP1 | 27,903 | 33.4929 | 14.6559 | 1.8438 | 89.9910 |
| Indep | 27,901 | 37.7412 | 5.6553 | 14.2900 | 100.0000 |

# 5.Empirical Results and Analysis

## 5.1Baseline Regression Analysis

Benchmark regression results reveal a significant positive correlation between corporate digital transformation and financial asset allocation. As shown in Table 3, a one-unit increase in digital transformation is associated with an average rise of 0.61 percentage points in short-term financial assets (fin1) and 0.14 percentage points in long-term holdings (fin2). This systemic alignment with information efficiency theory suggests that digital technologies enhance firms' ability to identify financial opportunities by reducing information costs and optimizing decision-making. The marginal effect on short-term assets is 4.5 times that on long-term assets, indicating that digital adoption particularly promotes liquid asset allocation—likely due to greater reliance on real-time information processing and reduced execution costs for high-frequency transactions through algorithmic trading and liquidity innovations.

Control variables further show that financial asset allocation is influenced by multiple factors. Firm Size exhibits dual effects, consistent with resource-based theories where larger firms prioritize long-term capital. The significantly negative coefficient of LEV underscores the role of financing constraints, as highly leveraged firms reduce financial investments due to liquidity pressures. ROA is positively correlated with short-term holdings, reflecting profitable firms' preference for liquid assets, but negatively correlated with long-term allocations, indicating risk aversion toward capital lock-in. The divergent effects of ListAge illustrate life-cycle dynamics: mature firms shift from short-term arbitrage to long-term strategic investment. While TOP1 is statistically significant, its economic magnitude is limited; Indep shows no significant effect, suggesting limited explanatory power of governance variables in financial asset allocation.

Although baseline regressions indicate a significant positive association, potential endogeneity may challenge causal interpretation. The following sections address this using propensity score matching and a staggered difference-in-differences design to strengthen identification.

Table 3: Baseline Regression Results

|  | (1)fin1 | (2)fin2 |
|---|---|---|
| Digital | 0.6104*** | 0.1364*** |
|  | (0.0310) | (0.0205) |
| Size | -0.0052*** | 0.0050*** |
|  | (0.0007) | (0.0004) |
| LEV | -0.1531*** | -0.0428*** |
|  | (0.0039) | (0.0025) |
| ROA | 0.1479*** | -0.0461*** |
|  | (0.0092) | (0.0057) |
| ListAge | -0.0196*** | 0.0282*** |
|  | (0.0011) | (0.0007) |
| TOP1 | 0.0004*** | -0.0001*** |
|  | (0.0001) | (0.0000) |
| Indep | 0.0001 | 0.0001 |
|  | (0.0001) | (0.0001) |
| Cons | 0.3884*** | -0.0962*** |
|  | (0.0148) | (0.0092) |

|         |           | R-squared |        | 0.1732 | 0.0877 |        |
|---------|-----------|-----------|--------|--------|--------|--------|
|         |           | N         |        | 27,895 | 27,875 |        |

Note: ***, **, * denote significance at the 1%, 5%, and 10% levels, respectively; standard errors in parentheses.

## 5.2 Staggered Difference-in-Differences

Building on the baseline correlation findings, this subsection addresses endogeneity using a PSM-DID framework, which involves two key steps: constructing comparable sample groups through propensity score matching (PSM) to ensure parallel trends, and implementing a multi-intensity multi-period staggered DID model to identify the net causal effect of digital transformation.

### 5.2.1 PSM Test

To satisfy the parallel trends assumption, firms are grouped by digital transformation intensity quantiles—treated (medium-high) and control (low)—and matched 1:3 using nearest-neighbor matching with a caliper of 0.2 based on model covariates. Post-matching, Table 4 shows no significant differences in covariates, passing balance tests and confirming comparability between groups. The kernel density plot (Fig.3-2) further indicates convergence in distribution between treated and control firms, supporting the common support assumption.

Table 4  Balance Statistics Before and After Matching

| Variable | Sample | Mean | | bias (%) | reduct |bias| (%) | t-test | |
|---|---|---|---|---|---|---|---|
| | | Treated | Control | | | t | p>|t| |
| Size | Unmatched | 22.334 | 22.139 | 15.200 | 95.200 | 11.880 | 0.000 |
| | Matched | 22.333 | 22.324 | 0.700 | | 0.680 | 0.494 |
| Lev | Unmatched | 0.421 | 0.436 | -6.700 | 94.600 | -5.290 | 0.000 |
| | matched | 0.420 | 0.419 | 0.400 | | 0.370 | 0.710 |
| ROA | Unmatched | 0.039 | 0.034 | 5.500 | 94.500 | 4.220 | 0.000 |
| | matched | 0.039 | 0.040 | -0.300 | | -0.300 | 0.764 |
| ListAge | Unmatched | 2.109 | 2.188 | -10.200 | 77.500 | -7.920 | 0.000 |
| | matched | 2.109 | 2.092 | 2.300 | | 2.170 | 0.030 |
| TobinQ | Unmatched | 2.062 | 2.077 | -1.100 | -51.500 | -0.830 | 0.405 |
| | matched | 2.061 | 2.037 | 1.600 | | 1.560 | 0.119 |
| TOP1 | Unmatched | 33.132 | 34.370 | -8.500 | 91.900 | -6.590 | 0.000 |
| | matched | 33.128 | 33.228 | -0.700 | | -0.660 | 0.507 |
| Indep | Unmatched | 37.933 | 37.335 | 10.700 | 94.000 | 8.260 | 0.000 |

|         | matched | 37.923 | 37.887 | 0.600 | 0.600 | 0.550 |

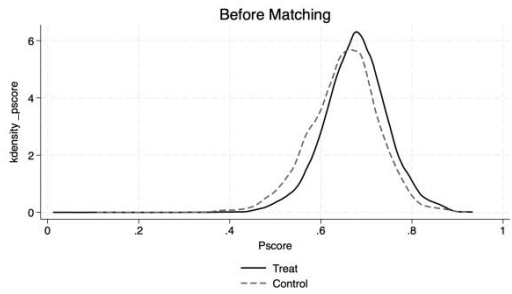 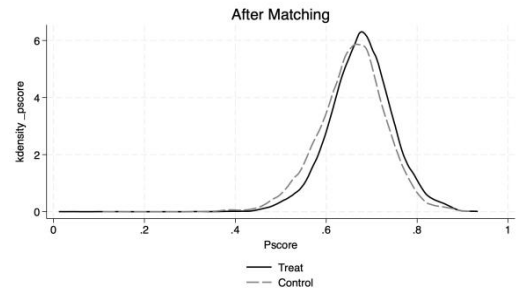

Fig.3-1 Kernel Density Estimate Before Matching    Fig.3-2 Kernel Density Estimate After Matching

### 5.2.2 DID Test Results

The staggered DID results in Table 5 indicate that the interaction term Digital×Post significantly promotes corporate financial asset allocation. Specifically, its coefficients are positive and significant at the 5% and 1% levels for fin1 and fin2, respectively. This supports digital transformation's triple-channel transmission mechanism: enhanced information processing reduces transaction costs for short-term assets; digitization of fixed assets unlocks capital for long-term equity investments; and improved data analytics strengthens the identification of high-return projects. Empirical evidence from Siemens AG (2015–2020) corroborates this—its Industrial 4.0 cash management system increased short-term financial assets from 4.8% to 7.1%[2], while its blockchain securitization platform raised long-term investments from 11.4% to 18.9%[3], aligning with the empirical estimates of 0.55% and 1.89% increases and highlighting the dual pathways of operational capital digitization (short-term) and fixed asset liquidity transformation (long-term).

Control variable analysis reveals deeper determinants of financial allocation. Digital transformation alters the role of firm Size: while it curbed short-term but promoted long-term allocation in traditional settings, it now enhances short-term holding through algorithmic trading systems. Meanwhile, the digitization of physical assets creates capital competition, turning the effect on long-term allocation negative. LEV shows a significantly negative correlation with short-term financial assets, indicating that highly leveraged firms reduce financial holdings due to financing constraints, yet it remains insignificant for long-term allocation. ListAge exhibits contrasting effects: negative for short-term allocation, reflecting mature firms' caution toward liquid investments, and positive for long-term allocation, suggesting greater competency in strategic investing. Indep remains insignificant, implying limited influence of corporate governance on financial asset allocation.

---

[2]SiemensAG.(2020) https://new.siemens.com/global/en/company/investors/financial-reports.html
[3]EuropeanCentralBank.(2021).https://doi.org/10.2866/91152

Table 5: Staggered DID Regression Results

|  | (1)fin1 | (2)fin2 |
|---|---|---|
| Digital×Post | 0.0055** | 0.0189*** |
|  | (0.0022) | (0.0012) |
| Size | 0.0164*** | -0.0082*** |
|  | (0.0033) | (0.0022) |
| LEV | -0.2016*** | -0.0067 |
|  | (0.0118) | (0.0073) |
| ListAge | -0.0868*** | 0.0352*** |
|  | (0.0055) | (0.0031) |
| Indep | -0.0002 | -0.0001 |
|  | (0.0002) | (0.0001) |
| Cons | 0.1014 | 0.1604*** |
|  | (0.0715) | (0.0480) |
| Firm fixed effects | YES | YES |
| Year fixed effects | YES | YES |
| R-squared | 0.6350 | 0.7520 |
| N | 21,159 | 21,159 |

Note: ***, **, * denote significance at the 1%, 5%, and 10% levels, respectively; standard errors in parentheses.

## 6.Endogeneity and Robustness Checks

## 6.1Addressing Endogeneity

To address potential endogeneity concerns such as reverse causality and omitted variables in the baseline regression, this study employs the density of long-distance optical cable lines at the provincial level (F_density) to construct an interaction-term instrumental variable (IV), F_density_post, for re-estimation. The IV satisfies both relevance and exclusion restrictions: provincial optical cable density reduces the marginal cost of corporate digital transformation, and its deployment—being government-led digital infrastructure—is exogenous to firms' financial decisions. By interacting F_density with a time indicator post within a difference-in-differences (DID) framework, the IV captures dynamic incentives for digital adoption after policy implementation while avoiding collinearity with firm fixed effects.

As shown in Table 6, the first-stage regression confirms a strong predictive power of the IV, with a KP Wald F-statistic of 1391.2, significantly exceeding the Stock-Yogo critical value of 16.38 at the 10% level, thus rejecting the weak instrument hypothesis.

The second-stage results indicate that digital transformation leads to a significant increase of 2.14 percentage points in short-term financial asset allocation and 1.71 percentage points in long-term allocation. This structural shift demonstrates that digitalization enhances firms' ability to allocate capital toward both liquid short-term instruments and long-term yield-generating assets through improved fund management.

Table 6: IV 2SLS Regression Results

| Variable | First Stage | Second Stage | |
|---|---|---|---|
| | Dep: did | Dep: fin1 | Dep: fin2 |
| $F\_density\_post$ | 0.0014*** | | |
| | (0.00003) | | |
| $Digital \times Post$ | | 0.0214*** | 0.0171*** |
| | | (0.0073) | (0.0038) |
| Size | 0.0534*** | 0.0144*** | -0.0084*** |
| | (0.0091) | (0.0033) | (0.0022) |
| LEV | -0.0617* | -0.1982*** | -0.0072 |
| | (0.0321) | (0.0114) | (0.0070) |
| ListAge | -0.0064 | -0.0868*** | 0.0360*** |
| | (0.0172) | (0.0055) | (0.0031) |
| Indep | -0.00174** | -0.00009 | -0.00012 |
| | (0.0009) | (0.0002) | (0.0002) |
| KP LM | 621.6340*** | 621.6340*** | 621.6340*** |
| KP Wald F | 1391.2130 | 1391.2130 | 1391.2130 |
| Stock-Yogo 10% critical value | 16.3800 | 16.3800 | 16.3800 |
| N | 21,777 | 21,777 | 21,777 |

Note: ***, **, * denote significance at the 1%, 5%, and 10% levels, respectively; standard errors in parentheses.

## 6.2 Robustness Checks

To ensure the reliability of the baseline results, robustness checks were conducted, including parallel trends, time trend, and placebo tests.

### 6.2.1 Parallel Trends Test

The validity of the difference-in-differences design hinges on the parallel trends assumption. Using an event study approach, we plot coefficient estimates for each year around the digital transformation policy implementation, omitting the year prior to the shock to avoid multicollinearity.

As shown in Figures 4-1 and 4-2, the estimated coefficients for both short-term (fin1) and long-term (fin2) financial assets are statistically indistinguishable from zero during the pre-treatment period, supporting the parallel trends assumption. In the first year following digital transformation, the coefficients become significantly positive and remain so thereafter. This sustained positive effect reflects how digitalization enhances financial asset allocation by improving liquidity management and optimizing investment structure, with a particularly persistent impact on long-term holdings.

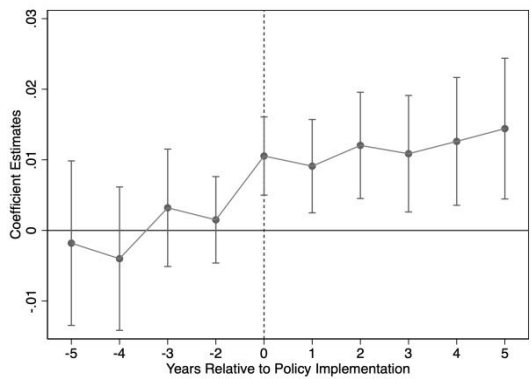
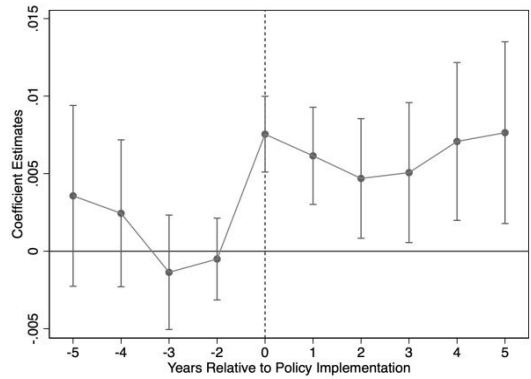

Fig.4-1 Parallel trend test results of fin1    Figure 4-2 Parallel trend test results of fin2

### 6.2.2 Time Trend Test

Although the parallel trends assumption is satisfied, unobserved heterogeneous time trends between treated and control groups may still exist. To address this, we incorporate firm-specific linear time trends into the model as follows:

$$Fin_{i,t} = \beta_0 + \beta_1(Digital_i \times Post_{i,t}) + \gamma X_{i,t} + \alpha_i + \delta_t + \lambda_0(1 - Digital_i) \times Trend_{i,t} + \varepsilon_{i,t}$$

Here, $Trend_{i,t}$ represents a firm-specific time trend, starting from the first year the firm appears in the sample and increasing by 1 annually. The term $(1 - Digital_i) \times Trend_{i,t}$ activates for control firms (where $Digital_i=0$), with $\lambda_0$ capturing the time trend coefficient for this group. The time trend for treated firms $\lambda_1 Digital_i \times Trend_{i,t}$ is omitted due to perfect collinearity with firm fixed effects $\alpha_i$[48].

Table 7 presents the regression results after controlling for firm-specific time trends.

The coefficient of Digital×Post remains positive and significant at the 5% and 1% levels for short-term and long-term financial assets, respectively, confirming that the positive effect of digital transformation is robust to time trends. The insignificant coefficients of the control group's time trend suggest no systematic pre-existing trends among low-digital firms. The treated group's trend, absorbed by firm fixed effects, indicates no additional heterogeneous trends among medium-high digital firms. These findings further validate the parallel trends assumption and strengthen the reliability of the main results.

Table 7: Time Trend Test Results

|  | (1)fin1 | (2)fin2 |
|---|---|---|
| Digital×Post | 0.0085** | 0.0173*** |
|  | (0.0035) | (0.0018) |
| Size | 0.0157*** | -0.0086*** |
|  | (0.0033) | (0.0022) |
| LEV | -0.1996*** | -0.0071 |
|  | (0.0114) | (0.0070) |
| ListAge | -0.0869*** | 0.0363*** |
|  | (0.0056) | (0.0031) |
| Indep | -0.0001 | -0.0001 |
|  | (0.0002) | (0.0002) |
| Cons | 0.1117 | 0.1689*** |
|  | (0.0709) | (0.0475) |
| (1-Digital)×Trend | 0.0007 | -0.0003 |
|  | (0.0005) | (0.0003) |
| Digital×Trend | (omitted) | (omitted) |
| Firm fixed effects | YES | YES |
| Year fixed effects | YES | YES |
| R-squared | 0.6307 | 0.7492 |
| N | 21,777 | 21,777 |

Note: ***, **, * denote significance at the 1%, 5%, and 10% levels, respectively; standard errors in parentheses.

### 6.2.3 Placebo Test

To rule out spurious correlations, a placebo test was conducted by randomly assigning the treatment status within the matched sample and generating a fictitious policy dummy. This process was repeated 3,000 times to obtain a distribution of placebo coefficients.

As shown in Figures 5-1 and 5-2, the estimated placebo coefficients for both fin1 and fin2 are centered around zero and approximately normally distributed. The actual estimated coefficients (0.0055 for fin1 and 0.0189 for fin2) fall in the extreme right tail of the placebo distributions, with placebo p-values < 0.001 (Table 8), indicating that the probability of obtaining such positive effects by chance is less than 1%. These results confirm that the positive impact of digital transformation on financial asset allocation is unlikely to be driven by unobserved random factors, lending robust support to a causal interpretation.

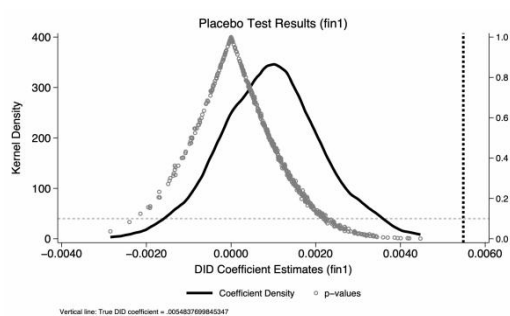
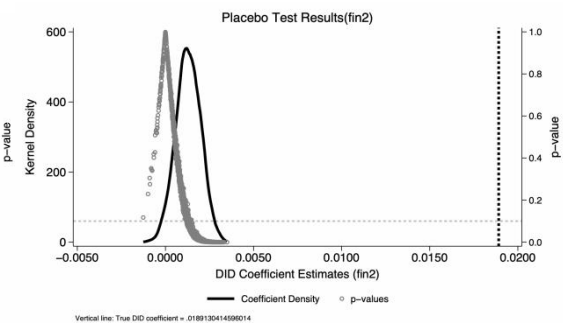

Fig.5-1 Parallel trend test results of fin1      Fig.5-2 Parallel trend test results of fin2

Table8 Placebo Test Statistics

| Metric | Value | |
| --- | --- | --- |
| | (1)fin1 | (2)fin2 |
| Actual DID Coefficient | 0.0055*** | 0.0189*** |
| Placebo Test p-value | 0.0000 | 0.0000 |
| True coefficient percentile | 100% | 100% |
| SD of Placebo Distribution | 0.0012 | 0.0007 |

Note: *** denotes significance at the 1% level

# 7.Heterogeneity Analysis

This section examines heterogeneity across three dimensions: firms' regional distribution, ownership type, and size. By identifying how digital transformation differentially influences financial asset allocation across firm characteristics, this study aims to reveal underlying patterns in technology dividend distribution and provide empirical support for targeted and context-sensitive digital economy policies.

## 7.1Heterogeneity Based on Region

Regional location significantly moderates the effect of digital transformation on financial asset allocation. Given China's interregional disparities in digital infrastructure and financial development, firms are grouped into eastern and non-eastern (central and western) regions based on registration addresses. Results in Columns (1) and (2) of Tables 9A and 9B show pronounced regional divergence. In short-term financial allocation, non-eastern firms exhibit stronger digital responsiveness, supported by a Chow test significant at the 1% level. This reflects poorer financial service coverage in these regions, where digital tools such as intelligent cash management systems help mobilize idle funds into highly liquid short-term instruments.

Conversely, eastern firms show a clear advantage in long-term financial asset allocation, with a Chow test significant at the 5% level. This is attributable to the region's mature financial ecosystem: well-developed venture capital networks facilitate access to high-quality projects, and blockchain-based equity platforms reduce investment risks, enabling firms to reinvest digitally unlocked capital into strategic long-term equity holdings.

This regional heterogeneity reveals a dual mechanism: in less developed financial regions, digital transformation mainly alleviates capital idleness and enhances short-term allocation, whereas in financially advanced regions, it synergizes with mature markets to promote long-term strategic investment. These findings imply that non-eastern regions should prioritize building provincial digital finance platforms, while eastern regions ought to promote deeper integration between digital technologies and capital markets.

## 7.2 Heterogeneity Based on Ownership

Ownership structure significantly moderates the effect of digital transformation on financial asset allocation. Using the nature of actual controllers, firms are categorized into state-owned enterprises (SOEs) and non-SOEs. As shown in Columns (3) and (4) of Tables 9A and 9B, SOEs exhibit a stronger response in short-term financial allocation, with a Chow test significant at the 1% level, attributable to dedicated policy support and stricter liquidity regulations that encourage high-liquid asset holdings. Conversely, non-SOEs show greater advantages in long-term allocation, also significant at the 1% level, reflecting their use of digital tools to overcome financing constraints and identify high-growth projects through supply chain finance platforms. This ownership-based divergence underscores how institutional settings shape digital outcomes: SOEs focus on short-term liquidity under regulatory pressure, while non-SOEs pursue long-term strategic gains. Policy implications include closer monitoring of SOE short-term investments and developing liquidity early-warning tools for non-SOEs to balance long-term strategy and financial stability.

## 7.3 Heterogeneity Based on Firm Size

Digital transformation's impact also varies by firm size. Firms are split into large and

SMEs based on the top 30% of asset size. Large firms show stronger effects in short-term allocation (Chow test p < 0.001), benefiting from resources to deploy digital treasury systems, AI arbitrage models, and scale-driven cash management. In contrast, SMEs excel in long-term allocation (significant at 1%), as cloud-based investment platforms lower entry barriers and smart tools mitigate analytical disadvantages, enabling access to private equity and strategic investments. Digital transformation thus helps SMEs overcome traditional size-related constraints while enhancing large firms' short-term operational efficiency.

Table9A: Short-term Financial Assets (fin1) Heterogeneity Test

| Variable | (1) Eastern region | (2) Non-eastern region | (3) SOEs | (4) Non-SOEs | (5) Large firms | (6) SMEs |
| --- | --- | --- | --- | --- | --- | --- |
| Digital×Post | 0.0096*** (0.0022) | 0.0141*** (0.0032) | 0.0221*** (0.0028) | 0.0051** (0.0023) | 0.0204*** (0.0030) | 0.0114*** (0.0022) |
| Chow test | 4.30 {0.0000***} | | 32.74 {0.0000***} | | 32.36 {0.0000***} | |
| Control Variables | YES | YES | YES | YES | YES | YES |
| Firm fixed effects | YES | YES | YES | YES | YES | YES |
| Year fixed effects | YES | YES | YES | YES | YES | YES |
| R-squared | 0.0385 | 0.0235 | 0.0228 | 0.0531 | 0.0243 | 0.0419 |
| N | 14,977 | 6,932 | 7,673 | 14,236 | 6,572 | 15,337 |

Note: ***, **, * denote significance at the 1%, 5%, and 10% levels, respectively; firm-level clustered standard errors in parentheses; {} contains empirical p-value for Chow test of group coefficient differences.

Table9B: Long-term Financial Assets (fin2) Heterogeneity Test

| Variable | (1) Eastern region | (2) Non-eastern region | (3) SOEs | (4) Non-SOEs | (5) Large firms | (6) SMEs |
| --- | --- | --- | --- | --- | --- | --- |
| Digital×Post | 0.0325*** (0.0014) | 0.0257*** (0.0022) | 0.0300* (0.0024) | 0.0321*** (0.0012) | 0.0271*** (0.0027) | 0.0306*** (0.0012) |
| Chow test | 2.23 {0.0053**} | | 18.69 {0.0000***} | | 13.15 {0.0000***} | |
| Control Variables | YES | YES | YES | YES | YES | YES |
| Firm fixed effects | YES | YES | YES | YES | YES | YES |
| Year fixed effects | YES | YES | YES | YES | YES | YES |
| R-squared | 0.0431 | 0.0298 | 0.0231 | 0.0531 | 0.0191 | 0.0476 |
| N | 14,977 | 6,932 | 7,673 | 14,236 | 6,572 | 15,337 |

Note: ***, **, * denote significance at the 1%, 5%, and 10% levels, respectively; firm-level clustered standard errors in parentheses; {} contains empirical p-value for Chow test of group coefficient differences.

# 8.Mechanism Tests

## 8.1Model Specification

While baseline and staggered DID analyses confirm that digital transformation promotes financial asset allocation, its underlying channels remain to be examined. To test the dual mechanisms of "investment channel broadening" and "information capacity enhancement" without incurring endogeneity bias from traditional mediation approaches, this study draws on Jiang (2022)[30] and examines two mediators: the ratio of long-term financial assets (Fin2_Ratio) and inefficient investment (IneffInvest). The model is specified as:

$$Mediator_{i,t} = \beta_0 + \beta_1(Digital_i \times Post_{i,t}) + \gamma X_{i,t} + \alpha_i + \delta_t + \varepsilon_{i,t} \quad (3)$$

where $Mediator_{i,t}$ represents either (1) Fin2_Ratio, capturing the optimization of asset maturity structure through broader investment channels (H1), or (2) IneffInvest, reflecting reduced capital misallocation via enhanced information processing (H2). $Digital_i \times Post_{i,t}$ is the DID treatment term, $X_{i,t}$ denotes control variables, and firm and year fixed effects ($\alpha_i, \delta_t$) are included.

### 8.2 Measurement of Mediator Variables

To accurately assess the mechanisms, the two mediators are constructed as follows:

Drawing on Peng et al. (2018), this study constructs Fin2_Ratio to capture the optimization of the maturity structure of financial assets. It is defined as the ratio of long-term financial assets to total financial assets[80]:

$$Fin2\_Ratio = \frac{fin2}{fin1+fin2} \quad (4)$$

where $fin1$ denotes short-term financial assets and $fin2$ represents long-term holdings. The ratio ranges between 0 and 1; a higher value indicates a greater preference for long-term assets, reflecting channel-broadening effects of digital transformation.

The residual measurement model of Richardson (2006) [41] is used to estimate the degree of inefficient investment. The model fits the optimal investment level of the enterprise through regression, and measures the degree of investment deviation with the absolute value of the residual. The model is set as follows:

$$Invest_{i,t} = \alpha_0 + \alpha_1 TobinQ_{i,t-1} + \alpha_2 LEV_{i,t-1} + \alpha_3 Cash_{i,t-1}$$
$$+ \alpha_4 Size_{i,t-1} + \alpha_5 ROA_{i,t-1} + \alpha_6 Age_{i,t-1} + \alpha_7 Invest_{i,t-1}$$
$$+ \delta_i + \delta_t + \varepsilon_{i,t} \quad (5)$$

Among them: $Invest$ is the current investment level, expressed as the ratio of cash paid for the construction of fixed assets, intangible assets and other long-term assets to the total assets of the previous period; $TobinQ$ represents investment opportunities in the previous period; financial leverage for the previous period; $Cash$ indicates cash flow held by the company, expressed as net cash flow from operating activities in the cash flow statement; $Size$ for the scale of the enterprise; $ROA$ is the net profit margin of total assets in the previous period; $Age$ is the age of market; $\delta_i + \delta_t$ denotes industry and annual dumb variables. In this paper, the model fitting value is taken as the expected investment level, and the absolute value of the regression residual is the degree of inefficient investment (IneffInvest), which indicates that the lower the investment efficiency and the more severe the capital allocation distortion.

### 8.3 Mechanism Test Results

Table 10, columns (3) and (4), reveal two core transmission mechanisms through which digital transformation influences financial asset allocation.

Column (3) shows a significant positive effect of digital transformation on the long-term financial asset ratio (Fin2_Ratio), confirming that digital technologies broaden firms' access to long-term investment instruments, consistent with H1. The adoption of tools such as blockchain-based securitization, cross-border REITs, and tokenized private equity expands firms' investment boundaries and optimizes the maturity structure of their assets. An increase in long-term assets not only helps capture intertemporal risk premiums but also reduces precautionary demand for liquidity through cash flow smoothing, thereby freeing up more capital for yield-generating assets. Begenau et al. (2018)[7] report that big data analytics improve the precision of long-term asset allocation by 41%. The BIS (2023)[6] further provides cross-country evidence that a one-standard-deviation (12%) increase in Fin2_Ratio leads to a 7.8% rise in long-term financial holdings, underscoring digital technology's profound restructuring effect on portfolio composition. The resulting improvement in the Sharpe ratio (Balyuk & Davydenko, 2024)[5] reflects enhanced risk diversification and return optimization.

Column (4) indicates that digital transformation significantly reduces inefficient investment (IneffInvest), supporting H2. By improving information acquisition and analytical capabilities, digital transformation helps identify and curtail negative-NPV investments, reallocating freed-up capital to efficient financial assets. Duchin et al. (2017)[20] find that a one-standard-deviation improvement in information processing capacity reduces inefficient investment by 13.4%, with 62% of the released capital converted into long-term financial assets (fin2), increasing fin2 allocation by 5.7 percentage points. This mechanism operates through what Gomber et al. (2017)[26] term a structural upgrade of the information ecosystem—using blockchain and IoT to integrate real-time, comprehensive data networks that break down information silos and support dynamically optimized investment decisions.

In summary, digital transformation affects corporate financial asset allocation through dual channels: "investment channel broadening" extends portfolio duration by expanding the supply of investable assets, while "information capacity enhancement" improves allocation efficiency on the demand side. Together, they shift corporate financial holdings from short-term liquidity management to long-term strategic investment, increasing the Sharpe ratio and strengthening capital resilience in uncertain environments.

Table 10: Two-Step Mediation Test Results

|           | (1) fin1    | (2) fin2    | (3) Fin2_Ratio | (4) IneffInvest |
|-----------|-------------|-------------|----------------|-----------------|
| did       | 0.005***    | 0.019***    | 0.084***       | -0.003***       |
|           | (0.002)     | (0.001)     | (0.003)        | (0.001)         |
| Size      | 0.013***    | -0.007***   | -0.024***      | -0.000          |
|           | (0.002)     | (0.001)     | (0.003)        | (0.001)         |
| LEV       | -0.176***   | -0.016***   | 0.100***       | 0.010***        |
|           | (0.009)     | (0.005)     | (0.013)        | (0.004)         |
| ROA       | 0.112***    | -0.037***   | -0.231***      | 0.050***        |
|           | (0.014)     | (0.008)     | (0.022)        | (0.009)         |
| ListAge   | -0.086***   | 0.036***    | 0.128***       | -0.010***       |
|           | (0.004)     | (0.002)     | (0.006)        | (0.001)         |
| TOP1      | 0.000       | 0.000       | -0.000         | -0.000***       |
|           | (0.000)     | (0.000)     | (0.000)        | (0.000)         |
| Indep     | -0.000      | -0.000      | 0.000          | 0.000           |
|           | (0.000)     | (0.000)     | (0.000)        | (0.000)         |
| TobinQ    | 0.001*      | -0.000      | -0.001         | 0.005***        |
|           | (0.001)     | (0.000)     | (0.001)        | (0.000)         |
| _cons     | 0.157***    | 0.138***    | 0.397***       | 0.063***        |
|           | (0.044)     | (0.026)     | (0.076)        | (0.013)         |
| R-squared | 0.582       | 0.742       | 0.650          | 0.019           |
| N         | 21414       | 21414       | 21418          | 19247           |

Standard errors in parentheses

# 9.Conclusion

## 9.1Research Findings

Using data from Chinese A-share listed companies (2010–2022), this study systematically examines the impact of digital transformation on corporate financial asset allocation and its underlying mechanisms. The main findings are as follows:

First, digital transformation significantly promotes corporate financial asset allocation overall, an effect that remains robust after addressing endogeneity and conducting rigorous checks. More importantly, its impact is heterogeneous across asset maturities. Both baseline regressions and causal identification show a stronger effect on short-term financial assets (fin1) than long-term holdings (fin2), consistent with information efficiency theory. Digital technologies enhance high-frequency, short-term financial transactions by improving liquidity management and reducing trading costs, thereby increasing holdings of monetary funds and trading financial assets.

Second, mechanism tests reveal that digital transformation operates through dual channels: "broadening investment channels" and "enhancing information capacity." On one hand, it breaks down information and access barriers, enabling firms to invest in long-term, high-yield instruments such as REITs, thus optimizing the maturity structure and risk-return profile of their portfolios. On the other hand, it significantly improves information acquisition, processing, and analytical capabilities, curbing inefficient investments (e.g., reckless physical expansion) and reallocating freed capital to high-efficiency financial assets.

Third, heterogeneity analyses show notable variations in digital effects. Regionally, the digital inclusion effect is more pronounced in non-eastern regions with weaker financial infrastructure, mainly enhancing short-term allocation, whereas eastern regions excel in long-term allocation due to their mature financial ecosystems. In terms of ownership and size, state-owned enterprises (SOEs) respond more strongly to short-term liquidity tools owing to regulatory constraints and resource advantages, while non-SOEs and SMEs are better able to convert digital dividends into long-term strategic investments, reflecting greater flexibility and risk tolerance.

## 9.2Strategies and suggestions

Based on the findings, this study proposes the following policy and managerial recommendations:

First, governments and regulators should adopt differentiated and targeted strategies for digital finance supervision and guidance. Given the pronounced effect of digital transformation on short-term financial allocation—which may entail arbitrage risks—and the varying long-term effects across regions and firm types, policy interventions should be more precise. Regulators ought to enhance real-time digital monitoring, especially of SOEs' short-term investments, using RegTech to establish liquidity risk early-warning systems and prevent systemic risks from financial disintermediation. Non-SOEs require stable policy expectations to sustain long-term investment incentives. Eastern regions and non-SOEs should be encouraged to align digital advantages with long-term strategic investments through tax incentives and industry funds, steering digital dividends toward innovation and industrial integration rather than short-term markets. Non-eastern regions need strengthened digital infrastructure and inclusive financial platforms to bridge the digital divide and expand long-term investment channels.

Second, firms ought to develop digital asset allocation strategies tailored to their specific characteristics. Large firms should leverage digital tools not only to improve short-term liquidity management but also to build AI-driven evaluation and risk control systems for long-term strategic investments, transforming informational and liquidity advantages into sustainable growth. SMEs and non-SOEs should use digital platforms to access advanced financial markets and focus on long-term holdings—such as supply chain finance and equity investments in innovative projects—that align with their core business, rather than pursuing short-term returns. Since digital transformation operates through dual channels—"information" and "access"—firms should treat it as a strategic imperative, not merely a technological upgrade. Investment in big data and AI can enhance decision-making and reduce inefficiency, while technologies like blockchain and robo-advisors can improve the structure and efficiency of long-term asset allocation.

Third, financial institutions and market participants should innovate in response to shifting corporate demand for diversified and multi-level digital asset allocation. Developing new digital finance products—such as ESG digital assets and supply chain finance solutions—and offering integrated, full-cycle financial services can support firms throughout digital transformation and foster healthier interactions between the real economy and financial markets.

## 9.3 Limitations and Future Research

Despite its rigorous approach, this study has several limitations that suggest fruitful future directions:

First, the sample is limited to A-share listed firms due to data availability. Future research could include unlisted or NEEQ-listed firms to improve generalizability. Further, while this study focuses on macro impacts, future work could examine the effects on specific firm types, such as "specialized, refined, unique, and innovative" enterprises or unicorn firms.

Second, the rapid evolution of digital technologies presents aproactive challenges. This study builds on current technologies like big data, cloud computing, and blockchain. However, disruptive innovations such as generative AI (AIGC) are reshaping information and decision-making patterns, potentially influencing corporate risk identification, asset pricing, and financial allocation in ways beyond existing theoretical frameworks. Continual tracking of technological advances and their economic consequences remains an essential and challenging research direction.

Third, macro-environment dynamics pose constraints. This study evaluates digital transformation within a relatively stable policy cycle, yet financial allocation decisions are influenced by macroeconomic fluctuations, industrial policies, and global financial conditions. Unforeseen regulatory shifts or global economic shocks may alter how digital transformation affects resource allocation. Although time trends are controlled, completely isolating these complex external dynamics is difficult. Future studies should adopt longer time frames and incorporate macro-institutional and international factors to better understand contextual influences.

# Appendix

## Bootstrap Mediation Test

Table 11 reports the bootstrap mediation test results (3,000 replications) to further validate the mediating roles of "investment channel broadening" and "information capacity enhancement" in the impact of digital transformation on corporate financial asset allocation. The results show that Fin2_Ratio (long-term financial asset ratio) plays a statistically significant mediating role in both pathways, providing strong support for H1. This confirms that digital transformation optimizes financial asset structure by broadening investment channels and increasing the share of long-term assets.

Inefficient investment exhibits a significant mediating effect only in the long-term allocation model, indicating that reduced investment inefficiency partially mediates the positive influence of digital transformation on long-term financial holdings, thus validating H2.

In summary, the bootstrap test confirms that digital transformation affects financial asset allocation through dual mechanisms: "investment channel broadening" and "information capacity enhancement." The increase in long-term financial assets serves as the core mediating channel, while the reduction in inefficient investment acts as a complementary mechanism in long-term allocation. The bootstrap approach effectively addresses endogeneity and low-power issues inherent in traditional stepwise regression, providing robust statistical evidence in support of both H1 and H2.

Table11 Bootstrap Mediation Test Results

|  | Effect | z | P>|z| | 95% Confidence Interval |
|---|---|---|---|---|
| Fin1-Fin2Ratio | Indirect | -19.85 | 0.000 | [-0.0295, -0.0242] |
|  | Direct | 15.43 | 0.000 | [0.0279, 0.0360] |
| Fin2-Fin2Ratio | Indirect | 18.98 | 0.000 | [0.0169, 0.0208] |
|  | Direct | -0.00 | 0.997 | [-0.0021, 0.0021] |
| Fin1-Inefficiency | Indirect | -0.01 | 0.991 | [-0.0001, 0.0001] |
|  | Direct | 2.56 | 0.010 | [0.0013, 0.0101] |
| Fin2-Inefficiency | Indirect | -2.27 | 0.023 | [-0.0001, -0.0000] |
|  | Direct | 14.91 | 0.000 | [0.0166, 0.0216] |